\documentstyle[emulateapj]{article}

\lefthead{Rey et al.}
\righthead{GALEX UV Observations of NGC 5128 Globular Clusters}

\begin{document}
\title{\large\textbf{Probing the Intermediate-Age Globular Clusters in NGC 5128 from Ultraviolet Observations}\normalsize\textnormal{}}

\author{Soo-Chang Rey\altaffilmark{1}, Sangmo T. Sohn\altaffilmark{2,3}, Michael A. Beasley\altaffilmark{4}, Young-Wook Lee\altaffilmark{2}, R. Michael Rich\altaffilmark{5}, Suk-Jin Yoon\altaffilmark{2}, Sukyoung K. Yi\altaffilmark{2}, Luciana Bianch\altaffilmark{6}, Yongbeom Kang\altaffilmark{1}, 
Kyeongsook Lee\altaffilmark{1}, Chul Chung\altaffilmark{2},
Tom A. Barlow\altaffilmark{3}, Karl Foster\altaffilmark{3}, Peter G. Friedman\altaffilmark{3}, D. Christopher Martin\altaffilmark{3}, Patrick Morrissey\altaffilmark{3}, Susan G. Neff\altaffilmark{7},
David Schiminovich\altaffilmark{8}, Mark Seibert\altaffilmark{9}, 
Ted K. Wyder\altaffilmark{3}, 
Jose Donas\altaffilmark{10}, Timothy M. Heckman\altaffilmark{6}, Barry F. Madore\altaffilmark{9}, Bruno Milliard\altaffilmark{10},
Alex S. Szalay\altaffilmark{6}, Barry Y. Welsh\altaffilmark{11}}

\altaffiltext{1}{Department of Astronomy and Space Science, 
Chungnam National University, Daejeon 305-764, Korea}
\altaffiltext{2}{Center for Space Astrophysics, Yonsei University, Seoul
120-749, Korea}
\altaffiltext{3}{California Institute of Technology, MC 405-47, 1200 East
California Boulevard, Pasadena, CA 91125}
\altaffiltext{4}{Instituto de Astrofisica de Canarias, Via Lactea, E-38200 La Laguna, Tenerife, Spain}
\altaffiltext{5}{Department of Physics and Astronomy, University of
California, Los Angeles, CA 90095}
\altaffiltext{6}{Department of Physics and Astronomy, The Johns Hopkins University, Homewood Campus, Baltimore, MD 21218}
\altaffiltext{7}{Laboratory for Astronomy and Solar Physics, NASA Goddard Space Flight Center, Greenbelt, MD 20771}
\altaffiltext{8} {Department of Astronomy, Columbia University, New York, NY 10027}
\altaffiltext{9}{Observatories of the Carnegie Institution of Washington, 813 Santa Barbara St., Pasadena, CA 91101}
\altaffiltext{10}{Laboratoire d'Astrophysique de Marseille, BP 8, Traverse du Siphon, 13376 Marseille Cedex 12, France}
\altaffiltext{11}{Space Sciences Laboratory, University of California at Berkeley, 601 Campbell Hall, Berkeley, CA 94720}

\begin{abstract}
We explore the age distribution of the globular cluster (GC) system of the nearby elliptical galaxy 
NGC 5128 using ultraviolet (UV) photometry from {\sl Galaxy Evolution Explorer} ({\sl GALEX}) observations, 
with UV$-$optical colors used as the age indicator. Most GCs in NGC 5128 follow the general trends of GCs 
in M31 and Milky Way in UV$-$optical color-color diagram, which indicates that the majority of GCs 
in NGC 5128 are old similar to the age range of old GCs in M31 and Milky Way. 
A large fraction of spectroscopically identified intermediate-age GC (IAGC) candidates 
with $\sim$ 3$-$8 Gyr are not detected in the FUV passband. Considering the nature of intermediate-age populations 
being faint in the far-UV (FUV) passband, we suggest that many of the spectroscopically identified IAGCs may be truly 
intermediate in age. This is in contrast to the case of M31 where a large fraction of spectroscopically 
suggested IAGCs are detected in FUV and therefore may not be genuine IAGCs but rather older GCs with developed blue horizontal branch stars.
Our UV photometry strengthens the results previously suggesting the presence of GC and stellar subpopulation 
with intermediate age in NGC 5128. The existence of IAGCs strongly indicates the occurrence of at least  
one more major star formation episode after a starburst at high redshift. 

\end{abstract}


\section{Introduction}
Globular cluster (GC) systems provide the signatures of formation and assembly 
histories of their host galaxies assuming that major star formations in galaxies are 
accompanied with global GC formation. 
Several scenarios have been proposed to account for the observational properties 
obtained for the GC systems (see a comprehensive review of Brodie \& Strader 2006).     
Many aspects of those scenarios are in favor of the currently accepted hierarchical 
galaxy formation theory (Press \& Schechter 1974) rather than the monolithic formation 
at high redshift (Eggen et al. 1962; Larson 1974).  
In this galaxy formation paradigm, constituent of galaxy mass including GCs is predicted 
to form through quiescent as well as merger/interaction-driven star formation 
(Kaviraj et al. 2007b).  

One of the best templates in the local universe for testing this scenario is the elliptical 
galaxy NGC 5128 due to its proximity. 
There have been several pieces of evidence supporting the picture that the 
NGC 5128 is the prototype for a postmerger elliptical galaxy (see Israel 1998 and references therein). 
Previous photometric and spectroscopic observations of GCs also suggest that merging and/or 
interaction events have played an important role in shaping its star cluster system 
(Peng, Ford, \& Freeman 2004a, b; Woodley et al. 2007; Beasley et al. 2008).

Constraining the formation scenario of the NGC 5128 GC system requires the understanding of
its global age distribution.  Clusters younger than the bulk of ancient Galactic counterparts 
are of particular interest because 
these objects represent the later stages of star 
formation histories in galaxies.  Recent spectroscopic observations suggest that 
NGC 5128 hosts a cluster population significantly younger than the old GCs in the Milky Way 
and M31 (Peng et al. 2004b).  Based on the spectroscopic observations for an increased sample of GCs, 
Beasley et al. (2008) reported the discovery of metal-rich, intermediate-age GCs (IAGCs) with 
ages of $\sim 3 - 8$ Gyr in NGC 5128.  They propose that this population may be the byproduct 
formed during merging events and/or interactions involving star formation and GC formation several 
gigayears ago.

However, it is important to note that age-dating of GCs via integrated spectra is hampered
by the degeneracy between age and the existence of hot old stellar population (e.g., blue 
horizontal branch [HB] stars) affecting the strength of age-sensitive line indices (Lee, Yoon, \& 
Lee 2000; Maraston et al. 2003; Thomas, Maraston, \& Bender 2003; Schiavon et al. 2004; 
Lee \& Worthey 2005; Trager et al. 2005; Cenarro et al. 2007). 
The effect of old blue HB stars in the integrated spectra can mimic young ages for old GCs,
raising a cause of concern that may cast doubt on the intermediate age nature of the GC in some galaxies. 

The UV colors (e.g. FUV$-V$ and FUV$-$NUV), on the other hand, are known to
provide robust age estimation of simple stellar populations (e.g., Yi 2003; Rey et al. 2005, 2007; 
Kaviraj et al. 2007a; Bianchi et al. 2007).  Kaviraj et al. (2007a) found that the age constraint is far superior when 
UV photometry is added to the optical colors and its quality is comparable or marginally better 
than the case of utilizing spectroscopic indices.

With the new approach using UV observations, in this {\it letter}, we take advantage of the combination of 
available optical photometry and the {\sl GALEX} ({\sl Galaxy Evolution Explorer}) UV photometry to 
confirm the existence of IAGCs and to explore the age distribution of the NGC 5128 GC system.
In the following sections, we emphasize the importance of the UV photometry as a probe of 
IAGCs in general. 
Comparing with GCs in M31 and the Milky Way with the aid of our population models,
we describe the overall age distribution of GCs and identification of IAGCs in NGC 5128. 
In this paper, we denote IAGCs as those having ages $\sim$ 3 $-$ 8 Gyrs.

\section{Observations and Data Analysis}

{\sl GALEX} (Martin et al. 2005) imaged one 1.25 deg circular field centered on 26 arcmin East and 
7 arcmin North of the NGC 5128 core in two UV bands: FUV (1350 -- 1750\AA) and NUV (1750 -- 2750\AA).
The images were obtained on April 2004, and are included in the 
{\sl GALEX} fourth and fifth data release (GR4/GR5)\footnote{http://galex.stsci.edu/gr4}.
Total integration times were 30,428 sec and 20,072 sec for NUV and FUV, respectively.
Preproccessing and calibrations were performed via the {\sl GALEX} pipeline
(Morrissey et al. 2005, 2007). {\sl GALEX} image has a sampling of 1.5 arcsec pixel$^{-1}$
which corresponds to 19 pc at the distance of NGC 5128 (3.9 Mpc, Woodley et al. 2007)  

Using the DAOPHOTII/ALLSTAR package (Stetson 1987), we performed aperture
photometry for all detected point sources in the {\sl GALEX} NGC 5128 field.
Aperture corrections were derived using moderately bright, isolated objects.
Flux calibrations were applied to bring all measurements into the
AB magnitude system (Oke 1990; Morrissey et al. 2005, 2007). 

Point sources in our {\sl GALEX} photometry were cross-matched using 
a matching radius of 3 arcsec with the 
catalog of Woodley et al. (2007). This catalog provides positions as well as 
optical magnitudes and mean radial velocities for 415 GCs in NGC 5128.
All spurious and ambiguous sources were rejected based on visual inspection.
The final sample of visually confirmed GCs are 157 and 35 in NUV and FUV, respectively.
We adopted a foreground reddening value of $E(B-V)$ = 0.11 for NGC 5128 (Woodley et al. 2007) and 
use the reddening law of Cardelli, Clayton, \& Mathis (1989). 
The full UV catalog and discussion of the UV properties of
GCs in NGC 5128 will be presented in a forthcoming paper.
Figure 1 shows the optical color-magnitude diagram (CMD) of GCs in NGC 5128 detected 
in the NUV and FUV bandpasses. For comparison,  
we overplot GCs in M31 detected from {\sl GALEX} observations (Rey et al. 2007). 
The CMD shows that most of the UV-detected objects in NGC 5128 and M31 have similar distribution 
and are confined to $V-I<1.05$.

\section{Ultraviolet as a Probe of Intermediate-Age Globular Clusters}

FUV flux plays an important role in identifying IAGCs.  Young ($< 1$ Gyr) stellar 
populations emit a substantial portion of their flux in the UV. 
Metal-poor old ($> 10$ Gyr) stellar populations also show large FUV to optical flux 
ratio due to the contribution of hot HB stars. 
On the contrary, intermediate-age ($\sim$ 3$-$8 Gyr) populations emit negligible 
amount of FUV flux since the constituent stars are not hot enough to produce a 
significantly large FUV flux (see Fig. 1 of Kaviraj et al. 2007a).
Consequently, if the IAGC candidates identified by spectroscopic observations are 
truly intermediate in age, they should be very faint or not detected in our {\sl GALEX} 
FUV photometry given our integration time and the detection limit (Lee \& Worthey 2005; 
Rey et al. 2007; Kaviraj et al. 2007a).

The first use of UV color as a tool for identifying IAGCs was demonstrated in our M31 study
(see Rey et al. 2007). Spectroscopic observations of M31 clusters have suggested the 
existence of IAGCs with mean age $\sim$ 5 Gyr (Burstein et al. 2004; Beasley et al. 2005;
Puzia et al. 2005).  However, based on {\sl GALEX} FUV detections of more than half of
M31 IAGC candidates, Rey et al. (2007) suggested that a large fraction of the spectroscopically 
identified IAGCs may not be truly intermediate in age but are rather old GCs with a developed 
blue HB sequence.  Among the 42 GCs in M31 whose ages are estimated by Kaviraj et al. 
(2007a), we find that four IAGC candidates turn out to be old GCs with $> 12$ Gyr.
By comparing of mass-to-light ratios of three IAGC candidates in M31 with those of old GCs, 
Strader et al. (2009) also found no evidence that M31 IAGC candidates 
are of intermediate in age.   

The most direct way to identify genuine IAGCs is to inspect CMDs of the clusters of interest. 
In the case of M31, $HST$ CMDs of two IAGC candidates B311 and B058 exhibit clearly 
developed blue HB sequences (Rich et al. 2005).  In a separate study, Chandar et al. (2006) 
showed that a star cluster in M33, C38, is a genuine IAGC with age $\sim 2$--5 Gyr based 
on the HST CMD and Balmer line measurements.  It is important to note that this cluster is 
also confirmed to be a genuine IAGC using the {\sl GALEX} FUV observations of M33 
(S. T. Sohn et al. 2009, in prep).  In any case, UV$-$optical color can be used to discriminate genuine 
IAGCs from the old GCs masquerading as IAGCs.

\section{Age Distribution of Globular Clusters in NGC 5128}

\subsection{Old Globular Clusters} 

Figure 2 shows the $V-I$ versus UV$-V$ diagrams.  We compare our NGC 5128 sample with those of the Milky Way 
(crosses, Sohn et al. 2006) and M31 (open circles, Rey et al. 2007) GCs whose age distributions 
are reasonably well constrained.  
We also show our simple stellar population (SSP) models constructed using the Yonsei Evolutionary 
Population Synthesis (YEPS) code (Lee, Yoon, \& Lee 2000; Lee et al. 2005; 
Rey et al. 2005, 2007; Yoon et al. 2006, 2008).

In Fig. 2, NGC 5128 GCs appear to show tight distribution around 12 Gyr model line similar to that of Milky Way, 
while GCs in M31 are rather scattered in $V-I$. This is partly due to the detection limit of optically
red GCs in NGC 5128 (see Fig. 1) and insufficient sample of Milky Way GCs obtained from previous UV observations of 
various satellites (see Sohn et al. 2006). Furthermore, Rey et al. (2007) reported the existence
of UV-bright metal-rich GCs with extreme hot blue HB stars in M31 (e.g., NGC 6388 and NGC 6441 in the Milky Way, Rich et al. 1997).
In this regard, some of the red ($V-I>1.0$) M31 GCs that show UV excess with respect to the 14 Gyr model line 
may be such peculiar objects. Considering these points, at a fixed $V-I$, 
the majority of GCs in three galaxies show similar spread in the UV$-V$ colors and are well 
accounted for by the 10--14 Gyr model lines.  This suggests that the mean age and age spread, 
at least, for old ($\geq 10$ Gyr) GCs are similar among GC systems of different galaxies, 
Milky Way, M31, and NGC 5128.

\subsection{Intermediate-Age Globular Clusters}

Beasley et al. (2008) found a population of intermediate-age 
and predominantly metal-rich ([Z/H] $> -1.0$) GCs (15 \% of the sample) from 
their spectroscopic observations.  Among the 21 IAGC candidates (age $\sim 3 - 8$ Gyr) 
identified by Beasley et al. (2008), we detect only two in the {\sl GALEX} FUV passband.

In Figure 3, we show the $V-I$ vs. $FUV-V$ diagram for the spectroscopically identified 
IAGC candidates in NGC 5128 (filled squares) and M31 (filled circles) detected in {\sl GALEX} 
FUV passband. Population model lines covering range of intermediate (3 and 8 Gyr) 
and old (10, 12, and 14 Gyr) ages are overplotted for guidance.
It is immediately apparent that all of the IAGC candidates of NGC 5128 and M31 detected 
in the FUV show similar distribution to those of old GCs with $> 10$ Gyr, {\it i.e.}, all FUV-detected 
IAGC candidates have significantly bluer $FUV-V$ colors than the 3 and 8 Gyr model lines.  
This indicates that IAGC candidates detected in the FUV are in fact old GCs 
($\ge 10$ Gyr) containing developed blue HB populations that contribute to the strong Balmer absorption lines.

It is important to note that, as shown in Fig. 3, most M31 IAGC candidates with $E(B-V)<0.16$ are 
detected in the {\sl GALEX} FUV (6 out of 7, see Rey et al. 2007 for the details).
If we restrict the sample of M31 IAGC candidates to match the observed optical brightness and 
color range ($M_{V}<-8$ and $V-I<1.05$, see Fig. 1) of the FUV-detected sample of NGC 5128 GCs, 
4 out of 5 M31 IAGC candidates are detected in FUV.
In the case of NGC 5128 GCs, only two out of 9 IAGC candidates are detected in the FUV.
Since all of the NGC 5128 GCs detected in the FUV cover similar range of $(FUV-V)_{o}$ colors of 
FUV-detected IAGC candidates in M31, most, if not all, spectroscopically identified IAGC candidates in 
NGC 5128 are not likely to be as bright as those in M31.

Among the 21 IAGC candidates identified by Beasley et al. (2008), 12 GCs are detected in the 
{\sl GALEX} NUV but not in the FUV. 
Whereas the FUV flux of old ($> 8$ Gyr) GC is almost entirely 
dominated by stars in the hot HB sequence, the NUV flux is influenced by both the HB stars 
and those on the main-sequence turnoff.
In this regard, we cannot rule out that some of the NUV-detected IAGC candidates are truly 
intermediate in age, despite the fact that NUV$-V$ is relatively insensitive to 
age variations compared to the FUV$-V$ (see Fig. 2).  To test this hypothesis, in Fig. 3, we show the bluer 
limits of the NUV-detected IAGC candidates having similar $V$ magnitudes of FUV-detected IAGCs.  
Most of the color limits are consistent with the NUV-detected IAGC candidates being $\sim 3 - 8$ Gyr in age.
In summary, our UV photometry suggests that NGC 5128 does possess a non-negligible 
fraction of IAGCs that are intrinsically faint in the FUV 
as proposed by previous spectroscopic studies.

\section{Discussion and Conclusions}

In this work, we explored the age distribution of GCs in the giant elliptical galaxy NGC 5128 using 
the UV colors. The majority of NGC 5128 GCs show age ranges similar to old GCs in M31 and 
the Galactic halo. Our most important result is that a large fraction of IAGCs identified by the 
spectroscopic observations are not detected in the {\sl GALEX} FUV passband and therefore may be 
truly intermediate in age.  This is in contrast to the case of M31 
GCs where the majority of IAGC candidates turned out to be old GCs with developed HB sequence 
based on their FUV$-V$ colors (see Rey et al. 2007). 

The existence of IAGCs in NGC 5128 supports the galaxy formation scenario accompanied with 
at least two major star formation episodes; e.g., hierarchical assembly of the protogalactic 
fragments or disks (Bekki et al. 2003; Beasley et al. 2002, 2003; Yi et al. 2004; Kaviraj et al. 2005).
In these models, some of the metal-rich GCs are formed from pre-enriched gas clouds and are on 
average younger than the metal-poor GCs.  Based on the kinematic analysis in combination with 
the age distribution of GCs, an alternative mechanism may have taken place where the NGC 5128 
formed its main body at early times and has gradually built up by minor mergers and gas-rich 
satellite accretions accompanied by star formation episodes (Woodley 2006; Woodley et al. 2007).

The presence of IAGCs in NGC 5128 has an interesting implication for the 
recent star formation (RSF) recently discovered using the large {\sl GALEX} UV sample of 
early-type galaxies at different redshifts ($0<z<1$; e.g., Yi et al. 2005; Kaviraj et al. 2007b, 
2008; Schawinski et al. 2007).  Kaviraj et al. (2008) found that high-redshift early-type galaxies 
in the range of $0.5<z<1$ exhibit typical RSFs in addition to the case of low-redshift 
($0<z<0.1$) early-type galaxies.  This provides a compelling evidence that RSFs in early-type 
galaxies are non-negligible over the last 8 billion years.  Furthermore, Kaviraj et al. (2008) suggest 
that up to 10$-$15\% of the mass of luminous ($-23<M_{V}<-20.5$) early-type galaxies such as 
NGC 5128 ($M_{V}=-21.08$, Gil de Paz et al. 2007) may have formed after $z=1$. 
These results imply that early-type galaxies in the local Universe are likely to possess 
intermediate-age stellar populations.  In this respect, IAGCs in NGC 5128 may be considered 
as relics of residual star formations that occurred during the last few billion years.

UV observations of the GC systems have been shown to provide important insights into the 
identification of IAGCs which is at present difficult to be identified solely by spectroscopic 
observations.  In particular, the Balmer line strengths themselves cannot reliably pin down the age of 
GCs because of the degeneracy between age and HB morphology.  FUV colors, on the other 
hand, can verify the contribution from hot stellar populations in GCs and help identify the 
true IAGCs.  Deep UV observations are highly anticipated for other galaxies with IAGC 
candidates identified by various spectroscopic and near-infrared photometric observations.

\acknowledgments
We thank Sugata Kaviraj for useful suggestions on the manuscript.
This work was supported by the Korea Research 
Foundation Grant funded by the Korean Government (MOEHRD) (KRF-2005-202-C00158) and 
the Korea Science and Engineering Foundation (KOSEF) 
through the Astrophysical Research Center for the Structure and Evolution of the Cosmos (ARCSEC).
{\sl GALEX} ({\sl Galaxy Evolution Explorer}) is a NASA Small Explorer, launched in 
April 2003. We gratefully acknowledge NASA's support for construction, 
operation, and science analysis for the {\sl GALEX} mission, developed in 
cooperation with the Centre National d'Etudes Spatiales of France and 
the Korean Ministry of Science and Technology. 


\clearpage

\begin{figure}
\epsscale{0.9}
\plotone{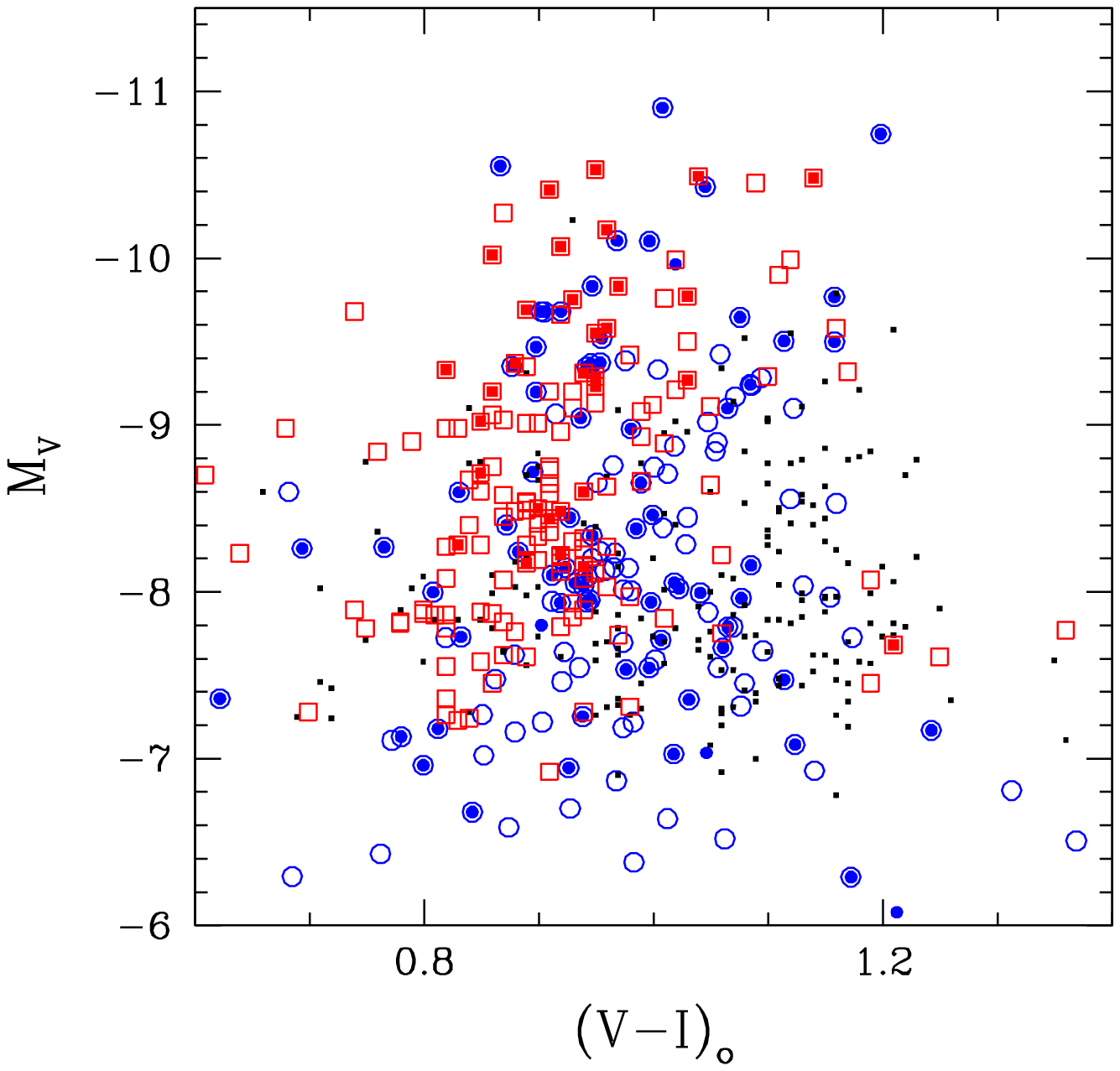}
\caption{$M_{V}$ vs. $(V-I)_{o}$ color-magnitude diagram of {\sl GALEX} UV-detected GCs in NGC 5128 (squares) 
and M31 (circles, Rey et al. 2007). Open and filled symbols are objects detected in NUV and FUV, respectively.
We note that all FUV-detected GCs in NGC 5128 are detected in NUV.
The small dots indicate GCs in NGC 5128 that are not detected in {\sl GALEX} UV observations.
          }
\label{f:f1}
\end{figure}

\begin{figure}
\epsscale{0.9}
\plotone{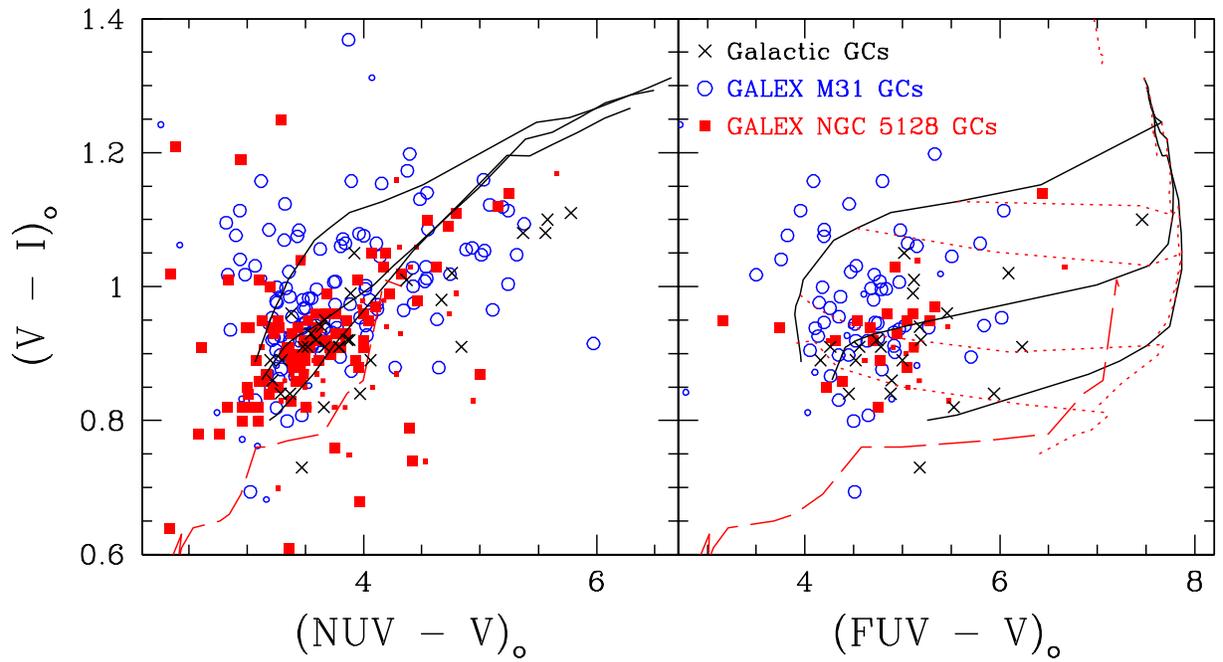}
\caption{$(V-I)_{o}$ vs. $(UV-V)_{o}$ diagrams of NGC 5128 (filled squares), 
Milky Way (crosses), and M31 (open circles) GCs. 
Large and small squares indicate NGC 5128 GCs with small and large magnitude errors in the UV passband 
(0.2 mag for NUV and 0.3 mag for FUV as the border line), respectively.
Large circles are M31 GCs with E$(B-V)<0.16$ from Barmby et al. (2000). Small circles are M31 GCs with no available reddening
information in Barmby et al., assuming that they are only affected by the Galactic foreground reddening of E$(B-V)$=0.10.   
We superpose our simple stellar population model lines with old (10, 12, and 14 Gyr; solid lines from bottom to top)
and young (long dashed line for 1 Gyr) ages. 
The dotted lines represent iso-metallicity lines varying from [Fe/H] = -2.0 to +0.5 dex (from bottom to top).
There is no significant
difference of distribution between red [$(V-I)_{o}$ $>$ 0.8] and old GCs in the three galaxies. 
}
\label{f:f1}
\end{figure}

\begin{figure}
\epsscale{0.9}
\plotone{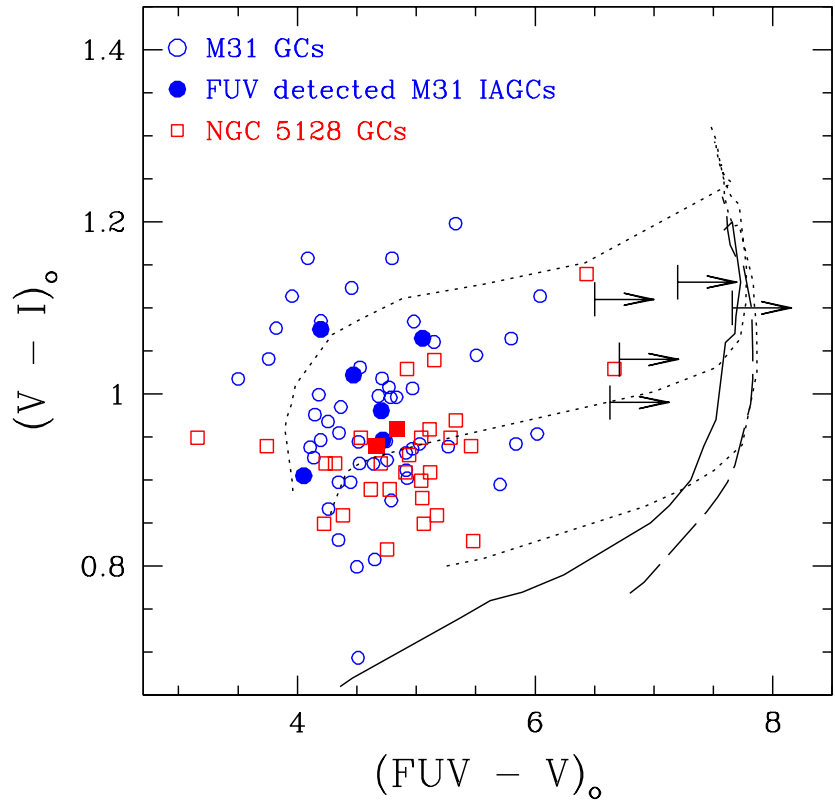}
\caption{$(V-I)_{o}$ vs. $(FUV-V)_{o}$ color-color diagram for the spectroscopically classified 
IAGC candidates in NGC 5128 (filled squares) and M31 (filled circles) detected in the {\sl GALEX} FUV passband. 
The model lines for intermediate ages (solid line for 3 Gyr and long dashed line for 8 Gyr) are overplotted  
in addition to the old (10, 12, and 14 Gyr; dotted lines from bottom to top) ones.   
All IAGC candidates of NGC 5128 and M31 detected in the FUV show similar distribution to that of 
old GCs (open circles and squares) with $>$ 10 Gyr. 
The color limit for each IAGC candidate of NGC 5128 not detected in the FUV is plotted with a vertical bar and a 
horizontal arrow pointing to the redder color.  Color limits were determined by adopting the flux of the faintest 
FUV-detected GC in NGC 5128.}
\label{f:f1}
\end{figure}

\end{document}